\documentclass[aps,12pt,showpacs,amsfonts,amsmath,amssymb,onecolumn]{revtex4}
\usepackage{mathrsfs}
\usepackage[small]{caption}
\usepackage{epsfig}
\usepackage{hyperref}
\usepackage{array}

\begin{document}

\rightline{ITP-UU-06/27, SPIN-06/23}

\title{\Large Antisymmetric Metric Field as Dark Matter}
\author{\Large Tomislav Prokopec}\email[]{t.prokopec@phys.uu.nl}
\author{Wessel Valkenburg}\email[]{w.valkenburg@phys.uu.nl}
\affiliation{Institute for Theoretical Physics (ITP) \& Spinoza
Institute,
             Minnaertgebouw, Leuvenlaan 4, Utrecht University
             3584 CE Utrecht, The Netherlands}
\date{\today}

\begin{abstract}
 We consider the generation and evolution of quantum fluctuations
of a massive nonsymmetric gravitational field ($B$-field) from inflationary
epoch to matter era in the simplest variant of the nonsymmetric theory of
gravitation (NGT), which consists of a gauge kinetic term and a mass term.
 We observe that quite generically a nonsymmetric metric field
with mass, $m_B\simeq  0.03(H_I/10^{13}~{\rm GeV})^4~{\rm eV}$, is a
good dark matter candidate, where $H_I$ denotes the inflationary
scale. The most prominent feature of this dark matter is a peak in
power at a comoving momentum, $k\sim \sqrt{m_B H_0}/(1+z_{\rm
eq})^{1/4}$, where $z_{\rm eq}$ is the redshift at equality. 
This scale corresponds roughly to the Earth-Sun distance.
%At the
%time of structure formation, $z\sim 10$, 
%this corresponds roughly to the Earth-Sun distance.
%one astronomical unit.
%length scale, $\sim 10^{8}~{\rm km}$, which is of the order of the
%Earth-Sun distance.
%While such a particle is consistent with all current short scale
%gravitational measurements, this dark matter may induce a modification to
%the gravitational force law at nanometer scale.

%2.2 10^11 m = 2.2 x 10^8 km 
%omega ^{-1} = 1.4 x 10^9 km 

\end{abstract}

\maketitle

\section{Introduction}

 Einstein's theory of relativity has passed all experimental tests
on laboratory and intermediate scales. On cosmological scales however,
in order to get a successful description of the Universe's dynamics
one requires addition of dark matter and dark energy, both of unknown
composition. An alternative is to extend Einstein's theory and hope
that a modified theory of gravitation would describe gravitation
on cosmological scale without a need for dark matter and/or dark energy.
Examples of extended theories of gravitation
which have both been used as alternatives 
to dark matter~\cite{Moffat:2004bm}
are the nonsymmetric theory of gravitation
(NGT)~\cite{Moffat:1978tr} 
and MOND~\cite{Milgrom:1983ca,Bekenstein:2004ne}.

 In Ref.~\cite{Prokopec:2005fb} we assumed that the antiymmetric
tensor field $B$ is dynamical,
and considered the cosmological evolution of quantum
fluctuations generated in de Sitter inflation of a massless
$B$-field, which is related by a duality transformation to the
Kalb-Ramond axion field. We have further shown that the evolution of
a massless $B$ field mimics that of a very light field. We then
presented some early results on the evolution of a massive
$B$-field, and observed that the scaling changes when the field
becomes nonrelativistic in radiation era. From our analysis of
quantum fluctuations of the physical mode of the antisymmetric
field, which corresponds to the longitudinal `magnetic' component,
it follows that the field couples conformally during de Sitter
inflation, such that the $B$-field correlator exhibits the spectrum 
of conformal vacuum fluctuations. This is contrary to what
has been claimed based on studies of the Kalb-Ramond axion, which
couples to gravitation just like a massless minimally coupled scalar
field, and therefore exhibits a nearly scale invariant spectrum
during inflation. In fact, because of the conformal coupling during
inflation, there cannot be a physical observable which exhibits a
scale invariant spectrum during inflation or subsequent epochs.
There is amplification however, which is induced by the matching at
the inflation-radiation transition, and based on which the spectral
amplitude gets enhanced at superhorizon scales with respect to the
conformal vacuum, but not enough to get a scale invariant
observable. Finally we observed that the spectral features of the
$B$ field fluctuations produced in inflation are similar to that of
gravitational waves, making it thus an alternative probe of
inflationary scale. There is an important difference in the spectrum
between the primordial gravitons and the $B$ field fluctuations: while
the primordial gravitons exhibit a scale invariant spectrum on
superhubble scales, the $B$-field fluctuations are suppressed.

 Building on an earlier work of Damour, Deser and
McCarthy~\cite{Damour:1992bt} and Clayton~\cite{Clayton:1996dz},
Janssen and Prokopec~\cite{Janssen:2006tj} have recently shown that there
is no nonsymmetric geometric theory which yields a dynamical $B$ field and
which is ``problem free,'' in the sense that the $B$-field dynamics is
ghost free and it is not marred by instabilities.
Furthermore the authors of~\cite{Janssen:2006tj}
have shown that the most general
(problem free) quadratic action in the $B$ field is of the form,
\begin{eqnarray}
S &=& S_{\rm HE} + S_{\rm B}
\,,\qquad S_{\rm B} = \int d^4 x {\cal L}_{\rm B}
\nonumber\\
 {\cal L}_{\rm B} &=&
          - \sqrt{-g}\frac{1}{12}H_{\mu\nu\rho} H^{\mu\nu\rho}
      - \sqrt{-g}\frac{1}{4}\Big(m_B^2 + \xi{\cal R}\Big) B_{\mu\nu}B^{\mu\nu}
\,,
\label{general lagrangian}
\end{eqnarray}
where
\begin{equation}
 H_{\mu\nu\rho} = \partial_\mu B_{\nu\rho}
                + \partial_\nu B_{\rho\mu}
                + \partial_\rho B_{\mu\nu}
\label{field strength H}
\end{equation}
is the antisymmetric field strength and
\begin{equation}
 S_{\rm HE} = \int d^4x {\cal L}_{\rm HE}
\,, \qquad
   {\cal L}_{\rm HE} = -\frac{1}{16\pi G_N}\Big({\cal R}+2\Lambda\Big)
\label{HE action}
\end{equation}
is the Hilbert-Einstein action, with $\Lambda$ being the cosmological
constant. In this paper we work in the units in which, $1/(16\pi G_N)=1$.
 From Eq.~(\ref{general lagrangian}) we see that the
coupling to the Ricci scalar ${\cal R}$
is the only allowed coupling to curvature.

 In order to simplify the problem further, we drop the coupling
to the Ricci scalar and set $\xi=0$ in Eq.~(\ref{general lagrangian}).
This term is expected to act as an additional mass term during inflation
era and thus introduce additional breaking of conformal invariance.
During radiation and matter era however, where the Ricci scalar is
either zero (radiation era) or of the order of the Hubble parameter
(matter era), we expect that the Ricci scalar term has no significant impact
on the evolution of the $B$-field.

 In a geometric theory the mass term is naturally induced by the
cosmological constant, in which case~\cite{Janssen:2006tj},
\begin{equation}
 m_B^2 \simeq 2\Lambda (1-2\rho + 8\sigma)
\,,
\label{mass vs Lambda}
\end{equation}
where $\rho$ and $\sigma$ are defined by the following decomposition of
the metric tensor,
\begin{equation}
  \bar g_{\mu\nu} =   g_{\mu\nu} + B_{\mu\nu}
                  + \rho B_\mu^{\;\alpha} B_{\;\alpha\nu}
                  + \sigma B^2 g_{\mu\nu} + {\cal O}(B_{\mu\nu}^3)
\,.
\label{metric decomposition}
\end{equation}
This is a general metric tensor decomposition in terms
of the symmetric metric tensor ($g_{\mu\nu}$) and antisymmetric metric
tensor ($B_{\mu\nu}$).
Yet there is no reason to assume that the mass term is fully
of geometric origin, and thus the relation~(\ref{mass vs Lambda}) needs not
to hold in general.

 Equation~(\ref{mass vs Lambda}) is also significant because
it implies that, if the $B$ field is of a geometric origin, then it
would be unnatural to assume that its mass vanishes. Indeed the
current observations, based on the luminosity-redshift relation of
distant supernovae Ia~\cite{Perlmutter:1998np,Riess:2004nr},
suggest that the cosmological term today is
of the order, $\Lambda \sim 10^{-84}~{\rm GeV}^2$. To maintain
generality we assume in this work that the mass $m_B$ is unspecified and
study its cosmological implications. In order to do that, we
canonically quantise the massive $B$ field in inflation, and follow
its subsequent dynamics in radiation and matter eras. Our main
finding is that the $B$ field with a mass of the order,
\begin{equation}
  m_B \simeq 0.03 \Big(\frac{10^{13}~{\rm GeV}}{H_I}\Big)^4~{\rm eV}
\end{equation}
which corresponds to a lengthscale,
$m_B^{-1}\simeq 7\times10^{-8}\big({H_I}/{10^{13}~{\rm GeV}}\big)^4~{\rm m}$,
is a good dark matter candidate.

 Since the $B$ field is produced in inflation and does not couple
to the matter fields, its spectrum is highly nonthermal. Indeed, we
find that the spectral power is peaked at a comoving momentum,
$k\simeq \sqrt{m_B H_0}/(1+z_{\rm eq})^{1/4}$, where $z_{\rm eq}$ is
the redshift at matter-radiation equality, and corresponds to a
physical scale at structure formation ($z\sim 10$), $k_{\rm
phys}^{-1}\sim 2\times 10^7~{\rm km}$. This peak is generated as a
consequence of a different nature of the vacuum states in inflation
and radiation era. This is the main feature by which this dark
matter can be distinguished from other dark matter candidates, which
typically obey a thermal statistic.

Another important feature are Fourier space pressure oscillations,
which occur after the second Hubble crossing.
 Although we find that the pressure of the $B$-field drops to zero
before the decoupling of the cosmic photon fluid, the Fourier
pressure components exhibit significant oscillations.
These oscillations may have a potentially observable impact on the
gravitational potentials, and they are thus the second distinct feature
of our dark matter candidate. The physical significance
of these spectral pressure oscillations should be further investigated.

\section{Stable linearised action}

 Taking the nonsymmetric gravity theory (NGT) as a starting point,
one can perform an expansion of some general covariant geometrical
action, in terms of antisymmetric perturbations on a symmetric
background metric. The (linearised) action then reads~\cite{Janssen:2006tj}
\begin{align}
S_{\rm NGT}=&\int d^4x \sqrt{-g}
\left\{\overline{R}-2\Lambda-\tfrac{1}{12}F_{\mu\nu\rho}F^{\mu\nu\rho}\right.\nonumber\\
&\left.+\left(\tfrac{1}{4}m_B^2+\theta_1 \overline{R}\right)
B_{\mu\nu}B^{\mu\nu}+\theta_2
\overline{R}_{\mu\alpha\nu\beta}B^{\mu\nu}B^{\alpha\beta}\right.\nonumber\\
&\left. + \theta_3
B^{\alpha}\/_{\rho}B^{\rho\beta}\overline{R}_{\alpha\beta}+\mathcal{O}(B^3)
\right\},\label{eq:action-effective}
\end{align}
where the parameters $\theta_1, \theta_2, \theta_3$ and $m_B$ are
defined by the initial action. We choose to work up to second order
in the nonsymmetric theory, and we will raise and lower indices with
the symmetric background FLRW metric,
\begin{align}
g_{\mu\nu}g^{\mu\rho}=&\delta^{\rho}_{\nu}, \\
g_{\mu\nu}B^{\mu\rho}=&B_{\mu}\/^{\rho}.
\end{align}

In Ref.~\cite{Damour:1992bt}, it was argued that NGT in general 
contains propagating ghosts or unacceptable constraints on dynamical degrees 
of freedom, when no cosmological term is present. For a simple
choice of the action, $\theta_n=m_B=0$ ($n=1,2,3$), no ghosts are found in
curved backgrounds~\cite{Prokopec:2005fb}. Nevertheless, 
when considered in a FLRW or Schwarzschild background,
the theory with $\theta_1\neq 0$ and/or $\theta_2\neq 0$ 
can develop instabilities even in the
presence of a cosmological term~\cite{Janssen:2006tj},
putting into question the geometric origin 
of the antisymmetric tensor field.
 In the light of these results, we choose $\theta_1=\theta_2=0$. In addition
for simplicity we choose $\theta_3=0$. 
The linearised action considered in this paper hence reads
\begin{align}
S_{\rm HE}+S_{\rm B}=&\int d^4x \sqrt{-g}
\left\{\overline{R}-2\Lambda-\tfrac{1}{12}F_{\mu\nu\rho}F^{\mu\nu\rho}
+\tfrac{1}{4}m_B^2 B_{\mu\nu}B^{\mu\nu}
\right\}.
\label{eq:action-effective-final}
\end{align}
In this equation we recognise the antisymmetric field strength of an
antisymmetric 2-form, analogous to the Kalb-Ramond
field~\cite{Kalb:1974yc},
\begin{align}
F_{\mu\nu\rho}&=\partial_\mu B_{\nu\rho}+\partial_\nu
B_{\rho\mu}+\partial_\rho B_{\mu\nu}\nonumber\\
&=\overline{\nabla}_\mu B_{\nu\rho}+\overline{\nabla}_\nu
B_{\rho\mu}+\overline{\nabla}_\rho B_{\mu\nu}.
\end{align}

\subsection{Field equations}
Following the action~\eqref{eq:action-effective-final}, the field
equations for $B_{\mu\nu}$ read
\begin{align}
a(\eta)^2\partial^\rho F_{\rho\mu\nu}-\frac{2a'}{a}F_{0\mu\nu}+m_B^2
a(\eta)^2B_{\mu\nu}=&0.\label{eq:eom2}
\end{align}
If we take the divergence of equation~\eqref{eq:eom2}, multiplied
with a factor of $a(\eta)^{-2}$, we are led to a self-consistency
condition,
\begin{align}
m_B^2 \partial^\mu B_{\mu\nu}=&0.\label{eq:gauge1}
\end{align}

Analogous to Maxwell theory, the antisymmetric tensor $B_{\mu\nu}$
can be decomposed in electric and magnetic components,
\begin{align}
B_{0i}\quad=-B_{i0}&=\quad E_i\label{eq:rewriting1}\\
B_{ij}\quad=-B_{ji}&=\quad-\epsilon_{ijk}B_k\label{eq:rewriting2}.
\end{align}
In this decomposition the field equations become
\begin{align}
E^L(x)=&0,\label{eq:eom-rw-start}\\
\left(\partial^2+m_B^2a(\eta)^2\right)\vec{E}^T(x)=&0,\\
\left(\partial^2-\frac{2a'(\eta)}{a(\eta)}\partial_0+m_B^2a(\eta)^2\right)B^L(x)=&0,\label{eq:eom-rw-bl}\\
\partial_0\vec{E^T}-\vec{\triangledown}\times\vec{B}^T(x)=&0.\label{eq:eom-rw-end}
\end{align}
Here we use superscripts $^T$ and $^L$ to denote the transverse and
longitudinal parts of the vectors $\vec{E}$ and $\vec{B}$.
Apparently the theory in this form possesses three degrees of
freedom. However, if the mass term disappears, the field equations
reduce to
\begin{align}
\partial^\rho a(\eta)^{-2} F_{\rho\mu\nu}=0.\label{eq:massless-f}
\end{align}
Equation~\eqref{eq:gauge1} no longer results as a self-consistency
condition. Though, the theory has gained a gauge freedom,
\begin{align}
B_{\mu\nu}&\rightarrow
B'_{\mu\nu}=B_{\mu\nu}+\partial_{\mu}\lambda-\partial_{\nu}\lambda,\nonumber\\
F_{\mu\nu\rho}&\rightarrow F'_{\mu\nu\rho}=F_{\mu\nu\rho},\nonumber
\end{align}
such that equation~\eqref{eq:gauge1} may be imposed as a choice of
gauge. Equation~\eqref{eq:massless-f} is then by definition solved
by
\begin{align}
a(\eta)^{-2}
F_{\rho\mu\nu}=\epsilon_{\rho\mu\nu\sigma}\partial^{\sigma}\phi.
\end{align}
Hence, the massless theory has only one (pseudoscalar) degree of freedom,
which is known as the Kalb-Ramond axion.
One could also have made an analysis of the degrees of freedom of this theory
by the means of its dual action, which at linearised level 
reduces to a massive Abelian gauge field action,
and therefore has three physical 
degrees of freedom~\cite{Valkenburg:2006}. However, in the presence of sources
the dual theory contains nonlocal source terms, justifying the
analysis in terms of the present variables. Note that the dual theory of
the massive antisymmetric tensor field should not be mistaken for
the massive Kalb-Ramond axion.

\subsection{Power spectrum}
What we eventually want to calculate is the spectrum of matter
density fluctuations, $\mathcal{P}_{\delta \rho}(\vec{k},\eta)$,
defined as
$\rho^{-2}_0\left<\delta\rho(\vec{x},\eta)\delta\rho(\vec{x}',\eta)\right>=\int
\frac{dk}{k}\,\mathcal{P}_{\delta\rho}$. In this case
$\mathcal{P}_{\delta\rho}=\frac{k^3}{2\pi^2}\left|\delta_{\vec{k}}\right|^2$,
and $\delta_{\vec{k}}$ is the fourier transform of $\frac{\delta
\rho}{\rho_0}$. $\rho_0$ denotes the average energy density and
$\left<\cdot\right>$ denotes the averaging procedure. In the case of
NGT, the power spectrum (i.e. the energy per decade in momentum
space) is given by
\begin{align}
\mathcal{P}_{\rm
B}(\vec{k},\eta)=\frac{k^3}{2\pi^2}T_{0}\/^{0}(\vec{k},\eta),\label{eq:p-vs-t-0-0}
\end{align}
where $T_{\mu\nu}$ is the stress-energy tensor of NGT,
\begin{align}
T_{\mu\nu}=\frac{2}{\sqrt{-g}}\frac{\delta}{\delta g_{\mu\nu}}S_{\rm
B}.
\end{align}
That is,
\begin{align}
T_{0}\/^{0}(\vec{x},\eta)=&\frac{1}{2
a(\eta)^{6}}\left\{\left(\partial_iB_j(\eta,\vec{x})\right)^2
%\right.\nonumber\\&\left.
+\left(\partial_0\vec{B}(\eta,\vec{x})+\nabla\times\vec{E}(\eta,\vec{x})\right)^2\right.\nonumber\\&\left.
\hskip 1.4cm
+a(\eta)^2m_B^2\left(\vec{E}(\eta,\vec{x})^2+\vec{B}(\eta,\vec{x})^2\right)\right\}.\label{eq:T-0-0}
\end{align}

\section{Canonical quantisation}
Due to the antisymmetry we find that the $0i$-components of the
canonical momenta are identically zero,
\begin{align}
\Pi^{0i}=\frac{\sqrt{-g}}{4}F^{00i}=0.
\end{align}
This residual gauge freedom is fixed by adding the Fermi
term~\cite{Kalb:1974yc},
\begin{align}
\mathcal{L}\rightarrow\mathcal{L}+\frac{\lambda
g_{\mu\nu}}{2}\left(\nabla_{\rho}B^{\rho\mu}\right)\nabla_{\sigma}B^{\sigma\nu},
\end{align}
such that the canonical commutation relations can still be imposed
in a covariant way. The Fermi term does not affect the longitudinal
magnetic component of the field.

In the next section we discuss the dynamics of the field during
inflation, radiation and matter era. One result is that all momenta
of interest, during inflation correspond to physical momenta which
are highly relativistic. Effectively the theory during inflation is
massless, hence the transverse degrees of freedom cannot obtain any
energy. From the classical field equations for the transverse
components, there is no indication that these degrees of freedom
should become important at any time. In the following we will
therefore neglect the transverse components.

In the context of quantising the theory with the Fermi term, it is
then sufficient to say that the theory can be covariantly
canonically quantised, and that for the component of interest we
have the commutation relation
\begin{align}
\left[\hat \Pi_{B_L}(x),\hat
B_{L}(y)\right]=-\frac{i}{2}\delta^{4}(x-y).
\end{align}

 Note that in Ref.~\cite{Janssen:2006tj} it is shown that in fact the
 transverse degrees of freedom contain the instabilities of the geometrical theory,
 involving couplings to the curvature tensors in the action~\eqref{eq:action-effective}. Canonical quantisation of transverse degrees of freedom
is carried out in Ref.~\cite{Valkenburg:2006}.

\section{Inflation, radiation and matter era}
We can perform a Fourier transformation,
\begin{align}
B^L(x)=&\int\frac{d^3k}{(2\pi)^{3/2}}\left\{e^{i\vec{k}\cdot\vec{x}}B^L\left(\eta,\vec{k}\right)a^{\phantom{\dag}}_{\vec{k}}\right.\nonumber\\
&\left.+e^{-i\vec{k}\cdot\vec{x}}B^{L \ast}\left(\eta,\vec{k}\right)a_{\vec{k}}^\dag\right\},\label{eq:Bmunu-fourier}\\
\end{align}
with the equal time relations
\begin{align}
\left[a^{\dag}_{\vec{k}},a^\dag_{{\vec{k}}'}\right]=&0,\label{eq:a-a-dagger1}\\
\left[a^{\phantom{\dag}}_{\vec{k}},a^{\phantom{\dag}}_{{\vec{k}}'}\right]=&0,\\
\left[a^{\phantom{\dag}}_{\vec{k}},a^\dag_{{\vec{k}}'}\right]=&
    \delta^3({\vec{k}}-{\vec{k}}')\label{eq:a-a-dagger2}.
\end{align}
The field equation for $B^L$ in momentum space becomes
\begin{align}
\left(\partial_0^2+k^2-\frac{2a'(\eta)}{a(\eta)}\partial_0+m_B^2a(\eta)^2\right)B^L(\eta,k)=&0,
\end{align}
Equations~\eqref{eq:a-a-dagger1}--\eqref{eq:a-a-dagger2} are
satisfied if the Wronskian satisfies
\begin{align}
W[B^L(\eta,k),B^{\ast
L}(\eta,k)]=&ia(\eta)^2,\label{eq:wronskian-for-l-mode}
\end{align}
We can conformally rescale $B^{L}$,
\begin{align}
\tilde{B}^L=\frac{1}{a(\eta)}B^L,
\end{align}
such that we have the field equation for $\tilde{B}^L$,
\begin{align}
\left(\partial_0^2+k^2+m_B^2a(\eta)^2+\frac{a''(\eta)a(\eta)-2a'(\eta)^2}{a(\eta)^2}\right)\tilde{B}^L=&0.
\end{align}
The expression for the power spectrum~\eqref{eq:p-vs-t-0-0}
containing only $B^L$ in momentum space becomes,
\begin{align}
\mathcal{P}_{\rm B}=\frac{k^3}{4\pi^2
a(\eta)^4}&\left\{\left|\partial_0 \tilde{B}^L (k,\eta)+\frac{a'}{a}\tilde{B}^L(k,\eta)\right|^2
%\right.\nonumber\\&\left.
+(k^2+a^2m_B^2)\left|\tilde{B}^{L}(k,\eta)\right|^2\right\}.\label{eq:p-long}
\end{align}

\subsection{De Sitter inflation}
Consider De Sitter inflation, in which the scale factor as a
function of conformal time $\eta$ is given by
\begin{align}
a(\eta)=\frac{-1}{H_I\eta}\quad\quad\mbox{(with
$-\infty<\eta<\frac{-1}{H_I}$)},
\end{align}
with $H_I$ the Hubble constant during inflation. The field equation
reads
\begin{align}
\left(\partial_0^2+k^2+\frac{m_B^2}{H_I^2\eta^2}\right)\tilde{B}^L_{inf}(\eta,k)=&0,\label{eq:qeom-rw-BLtilde}
\end{align}
solved by
\begin{align}
\tilde{B}^L_{inf}(\eta,k)=\alpha_B^{inf}\sqrt{\eta}
Z^{(1)}_{\frac{1}{2}\sqrt{1-(4m_B^2/H_I^2)}}(k\eta)+\beta_B^{inf}\sqrt{\eta}
Z^{(2)}_{\frac{1}{2}\sqrt{1-(4m_B^2/H_I^2)}}(k\eta).\label{eq:BL-inf}
\end{align}
Here $Z^{(n)}_{\nu}$ denotes any general pair of Bessel
functions~\cite{1994tisp.book:G} that forms a basis for the solution
space. For every momentum $k$ there has been an $\eta$ during
inflation such that $k_{\rm phys}>m_B a(\eta)$. As during inflation
the effectively massless $B^L$ field is conformally invariant, its
vacuum state is given by the conformal
vacuum~\cite{Birrell:1982ix,Bunch:1978yq} at early time (as $\eta
\rightarrow - \infty$). The vacuum state mode function reads,
\begin{align}
\tilde
B^L_{inf}(\eta,k)&=-i\sqrt{\frac{\pi\eta}{4}}
  H^{(2)}_{\frac{1}{2}\sqrt{1-(4m_B^2/H_I^2)}}(k\eta)
.\label{eq:BL-inf-normd}
\end{align}
Note that in the limit when $m_B\rightarrow 0$, 
the field is in conformal vacuum.
This is the relevant limit, since we are interested in the physical momenta
which at the end of inflation satisfy the relation, 
$H_I\gg k/a(\eta_{end})\gg m_B$. Indeed, the physical momenta that are within
the Hubble radius today (and of course smaller than the comoving $H_I$)
all satisfy this condition. 
Moreover, since conformal invariance is broken by 
the stress energy tensor~(\ref{eq:T-0-0}), from the point of view 
of energy density, superhorizon modes in fact undergo amplification
during inflation. This is in agreement with the stress
energy tensor of the massless Kalb-Ramond axion, which to a good approximation
corresponds to the dual description during inflation.

\subsection{Radiation era}

We choose the simple model of a sudden transition from inflation era
to radiation era, with a scale factor
\begin{align}
a(\eta)=H_I\eta\quad\quad\mbox{with
$(\frac{1}{H_I}<\eta<\eta_e)$}.\label{eq:scalefactor-rad-era}
\end{align}
Conformal time at the time of radiation-matter equality is denoted
by $\eta_e$.

The field equation for the longitudinal field in this era is given
by
\begin{align}
\left(\partial_0^2+k^2-\frac{2}{\eta}\partial_0+m_B^2H_I^2\eta^2\right)B^L_{\rm
rad}(\eta,k)=0.\label{eq:eom-bl-rad}
\end{align}
This field equation is in general solved by a special form of the
confluent hypergeometric function, namely the Whittaker function
$M_{\lambda,\mu}(x)$~\cite{1960slater}. We prefer to choose a
particular linear combination of Whittaker functions,
\begin{align}
B^L_{\rm rad}(\eta,k)=&\alpha_{B}^{\rm
rad}(k)\tilde{M}^{(1)}\left[\frac{3}{4},\frac{k^2}{4m_B
H_I};k
\eta\right]
%\nonumber\\&
+\beta_{B}^{\rm
rad}(k)\tilde{M}^{(2)}\left[\frac{3}{4},\frac{k^2}{4m_B
H_I};k \eta\right],
\label{BL-rad}
\end{align}
with
\begin{align}
\tilde{M}^{(1)/(2)}\Big[\frac34,\zeta;k \eta\Big]=&e^{\frac{i
\pi}{4}}\sqrt{\frac{\pi H_I \eta}{4 m_B}}
\left[\frac{\zeta^{-\frac{3}{4}}}{\Gamma\left(\frac{5}{2}\right)}M_{i\zeta,\frac{3}{4}}\left(-i\frac{
k^2\eta^2}{4\zeta}\right)
%\right.\nonumber\\& \left.
\pm
i\frac{\zeta^{\frac{3}{4}}}{\Gamma\left(-\frac{1}{2}\right)}M_{i\zeta,-\frac{3}{4}}\left(-i\frac{
k^2\eta^2}{4\zeta}\right)\right],\label{eq:whittaker-hankel1}%\\
%\tilde{M}_{\frac{3}{4}}^{(2)}[\zeta;k \eta]&=&e^{\frac{i
%\pi}{4}}\sqrt{\frac{\pi H_I }{4 m_B
%}}\left[\frac{\zeta^{-\frac{3}{4}}}{\Gamma\left(\frac{5}{2}\right)}M_{i\zeta,\frac{3}{4}}\left(-i\frac{
%k^2\eta^2}{4\zeta}\right)-i\frac{\zeta^{\frac{3}{4}}}{\Gamma\left(-\frac{1}{2}\right)}M_{i\zeta,-\frac{3}{4}}\left(-i\frac{
%k^2\eta^2}{4\zeta}\right)\right]\label{eq:whittaker-hankel2},
\end{align}
because this tends to the asymptotic Bunch-Davies vacuum (i.e. the
massless limit as $\eta \rightarrow 0$),
\begin{align}
B^L_{\rm rad}(\eta,k)=&\alpha_{B}^{\rm
rad}(k)\eta^{\frac{3}{2}}H^{(1)}_{\frac{3}{2}}(k
\eta)+\beta_{B}^{\rm rad}(k)
\eta^{\frac{3}{2}}H^{(2)}_{\frac{3}{2}}(k
\eta)\nonumber\\
&+\mathcal{O}\left(\frac{m_B^2}{H_I^2}\right).
\end{align}
The Hankel functions are given by
\begin{align}
\eta^{\frac{3}{2}}H^{(1)}_{\frac{3}{2}}(k \eta)&=
\sqrt{\frac{2}{\pi k^3}}\left\{-i -k\eta\right\}e^{i k \eta},\nonumber\\
\eta^{\frac{3}{2}}H^{(2)}_{\frac{3}{2}}(k \eta)&= \sqrt{\frac{2}{\pi
k^3}}\left\{i -k\eta\right\}e^{-i k \eta}.\nonumber
\end{align}
With this choice of solution, the canonical condition,
equation~\eqref{eq:wronskian-for-l-mode} is satisfied if
\begin{eqnarray}
W[B^L_{\rm rad}(\eta,k),B^{L\ast}_{\rm
rad}(\eta,k)]=\left(|\beta_{B}^{\rm rad}|^2-|\alpha_{B}^{\rm
rad}|^2\right)i a(\eta)^2.\label{eq:wronskian-b-bst-rad}
\end{eqnarray}
The coefficients $\alpha$ and $\beta$ in the solution are to be
defined by a continuous matching of $B^L$ and its derivative at the
time of transition from inflation to radiation era.  For notational simplicity,
we write $\tilde M^{(1)}$ and $\tilde M^{(2)}$ for the fundamental solutions
in radiation era~(\ref{BL-rad}), 
and $\tilde B^L_{inf}$ for the mode function~(\ref{eq:BL-inf-normd}) 
during inflation all evaluated at the inflation-radiation transition. This 
then leads to
\begin{eqnarray}
\alpha_{B}^{\rm rad}&=&\frac{1}{\tilde M^{(1)}{\tilde M^{(2)\prime}}
                       - \tilde M^{(2)}{\tilde M^{(1)\prime}}}
                        \left\{\tilde B^L_{inf}{\tilde M^{(2)\prime}}
                             -\tilde M^{(2)}\tilde B^{L\,\prime}_{inf}\right\}
\nonumber\\
&=&\frac{1}{W\left[\tilde M^{(1)},\tilde M^{(2)}\right]}
                  \left\{\tilde B^L_{inf}{\tilde M^{(2)\prime}}
                 -\tilde M^{(2)}\tilde B^{L\,\prime}_{inf}
              \right\}
\nonumber\\
&=&-i  %\sqrt{\frac{\pi}{4H_I}}
       \left\{\tilde B^L_{inf}{\tilde M^{(2)\prime}}
              -\tilde M^{(2)}\tilde B^{L\,\prime}_{inf}\right\}.
\end{eqnarray}
Likewise we find
\begin{equation}
\beta_{B}^{\rm rad}= i %\sqrt{\frac{\pi}{4H_I}}
                     \left\{\tilde B^L_{inf}\tilde M^{(1)\prime}
                   - \tilde M^{(1)}\tilde B^{L\,\prime}_{inf}\right\}
.
\end{equation}
Explicitly, that is
\begin{align}
\alpha_{B}^{\rm rad}&=-\frac{1}{2}\frac{H_I^2}{k^2}+\mathcal{O}\left(\frac{m_B}{H_I}\right)\label{eq:a-rad} \\
\beta_{B}^{\rm rad}&=-\frac{1}{2}\frac{H_I^2}{k^2} \left[1-2 i
\frac{k}{H_I} - 2 \frac{k^2}{H_I^2}\right]e^{2 i
\frac{k}{H_I}}+\mathcal{O}\left(\frac{m_B}{H_I}\right).\label{eq:b-rad}
\end{align}
Thus we can see that for superhubble modes at the end of inflation,
\begin{align}
\beta_{B}^{\rm rad}\simeq\alpha_{B}^{\rm rad}
  =- \frac{1}{2}\frac{H_I^2}{k^2}
    \Big[1+ \mathcal{O}\left(\frac{m_B}{H_I},\frac{k}{H_I}\right)\Big]
.
\end{align}
This Bogoliubov mixing~\cite{Bogoliubov:1958aa,Birrell:1982ix} of the modes is 
caused by the mismatch of the vacua during inflation and radiation era,
and loosely speaking can be thought of as particle production
at the inflation-radiation transition. 

%Apparently, the vacuum states of the longitudinal field during
%inflation and radiation era mismatch. Matching the vacuum state at
%the end of inflation continuously to the radiation era solution
%causes a mixing of modes during the radiation era, exciting the
%field. This is in fact a Bogoliubov
%transformation~\cite{Bogoliubov:1958aa}. The transformation is
%explicitly shown as both matching coefficients are nonzero, where
%the basis of solution space is aligned to the asymptotic vacuum,
%equation~\eqref{eq:whittaker-hankel1}%~and~\eqref{eq:whittaker-hankel2},
%such that the vacuum state is represented by only one nonzero
%coefficient.

%This process comes in addition to the antidamping of superhubble
%modes. The horizon crossing extends the lifetime of fluctuations,
%while the mismatching actually brings the field in an excited state.

\subsection{Matter era}
During matter era the scale factor is given by
\begin{align}
a(\eta)=\frac{H_I}{4\eta_e}(\eta+\eta_e)^2.
\end{align}The field equation for $\tilde{B}^L$ during matter era
reads
\begin{align}
\left(\partial_0^2+k^2-\frac{6}{\left(
\eta+\eta_e\right)^2}+m_B^2\frac{H_I^2}{16\eta_e^2}(\eta+\eta_e)^4\right)\tilde{B}^L_{\rm
mat}(\eta,k)=0.
\end{align}
As the field equation contains a term $\propto \eta^4$, it cannot be
solved analytically. However, for different values of $k\eta$ the
field equation effectively takes various forms, shown in
tabel~\ref{table:eff-fieldeq-matt-era}.

{\Large
\begin{table*}

%{||l|rl|rl||}

 \begin{center}
\renewcommand{\arraystretch}{1.6}
 \begin{tabular}{||l|rl|rl||}\hline
 Region&\phantom{{\large A}} Effective field equation& & Scale& \\ \hline \hline
$\,I $ & $\left(\partial_X^2+1-\frac{6}{X^2}\right)\tilde{B}^L_{\rm
mat}(X)$&$=0\quad$& $\quad X^2\quad\ll$&$\frac{4k\eta_{e}}{\xi}$
 \\ \hline
 $\,II$  & $\left(\partial_X^2+1-\frac{6}{X^2}+\frac{\xi^2}{16 k \eta_e}X^4\right)\tilde{B}^L_{\rm mat}(X)$&$=0\quad$& $\quad X^2\quad\simeq$&$\frac{4k\eta_{e}}{\xi}$%\\&&&&
 \\ \hline
 $\,III$& $\left(\partial_X^2+\frac{\xi^2}{16 k^2 \eta_e^2}X^4\right)\tilde{B}^L_{\rm mat}(X)$&$=0\quad$&
 $\quad X^2\quad\gg$&$\frac{4k\eta_{e}}{\xi}$\, and \, $X^3\gg \frac{4 \sqrt{6} k \eta_e}{\xi}$\\ \hline
 \end{tabular}\end{center} \caption{The effective field equations for different values of $X\equiv k\tilde{\eta}\equiv k (\eta_e+\eta)$. The expressions are written
  in terms of the parameter $\xi\equiv\frac{m_B
H_I}{k^2}$.}\label{table:eff-fieldeq-matt-era}
\end{table*}
}

 In region $I$ the solution is given by
\begin{align}
\tilde{B}^L_{\rm mat}(\tilde{\eta},k)=\alpha_{B}^{\rm mat}
\tilde{\eta}^{\frac{1}{2}}H_{\frac{5}{2}}^{(1)}(k\tilde{\eta})+\beta_{B}^{\rm
mat} \tilde{\eta}^{\frac{1}{2}}H_{\frac{5}{2}}^{(2)}(k\tilde{\eta}),
\end{align}
with again the coefficients given by a continuous matching at the
radiation-matter equality,
\begin{align}
\alpha_{B}^{\rm mat}=&\alpha_{B}^{\rm rad}\left(4i +
\frac{2}{k\eta_e}-\frac{i}{2 k^2
\eta_e^2}\right)e^{-ik\eta_e}
%\nonumber\\&
+\beta_{B}^{\rm rad}\frac{i}{2k^2\eta_e^2}e^{-3ik\eta_e}
,
\label{eq:a-mat-massless}\\
\beta_{B}^{\rm mat}=&\alpha_{B}^{\rm rad}\frac{-i}{2k^2\eta_e^2}e^{3ik\eta_e}
%\nonumber\\&
+\beta_{B}^{\rm rad}\left(-4i + \frac{2}{k\eta_e}+\frac{i}{2 k^2
\eta_e^2}\right)e^{ik\eta_e}\label{b-mat-massless}.
\end{align}
In region $II$ the solution of the field equation has to be solved
numerically. In region $III$, the field equation reduces to
\begin{align}
\left(\partial_{\tilde{\eta}}^2+\omega(\eta)^2\right)\tilde{B}^L_{\rm
mat}(\tilde{\eta},k)=0.
\end{align}
If $\frac{\omega'}{\omega}<\omega$ and
$\frac{\omega''}{\omega}<\omega^2$, which is the case for the
demanded values of $k\tilde{\eta}$ in this region, the solution can
be approximated by
\begin{align}
\tilde{B}^L_{\rm mat}(\tilde{\eta},k)
  = \frac{\alpha_{B}^{\rm mat}}{\sqrt{2 \omega(\tilde{\eta}) }}
       {\rm e}^{i \int^{\tilde{\eta}} d\eta \,\omega (\eta) }
   +\frac{\beta_{B}^{\rm mat}}{\sqrt{2 \omega(\tilde{\eta}) }}
  {\rm e}^{-i \int^{\tilde{\eta}} d\eta \, \omega (\eta) },
\end{align}
obeying the condition
\begin{align}
W[u_1,u_1^*]\simeq - i,\qquad
   u_1 = \frac{1}{\sqrt{2 \omega(\tilde{\eta}) }}\,
         {\rm e}^{i \int^{\tilde{\eta}} d\eta \omega (\eta) }
\,.
\end{align}
The explicit expressions for $\alpha$ and $\beta$ are rather
complicated and for simplicity omitted.

\section{The spectrum}
The power spectrum of the longitudinal mode,
equation~\eqref{eq:p-long}, in momentum space reads
\begin{align}
\mathcal{P}_{\rm B}(\eta,k)=&\frac{k^3}{4\pi^2
a(\eta)^{4}}\left\{\left|\partial_0\tilde{B}^L(\eta,k)+\frac{a(\eta)'}{a(\eta)}\tilde{B}^L(\eta,k)\right|^2\right.\nonumber\\
&\left.+(k^2+a(\eta)^2m_B^2)\left|\tilde{B}^L(\eta,k)\right|^2\right\}.\label{eq:powerspectrum}
\end{align}

\begin{figure*}[t]
\epsfig{file=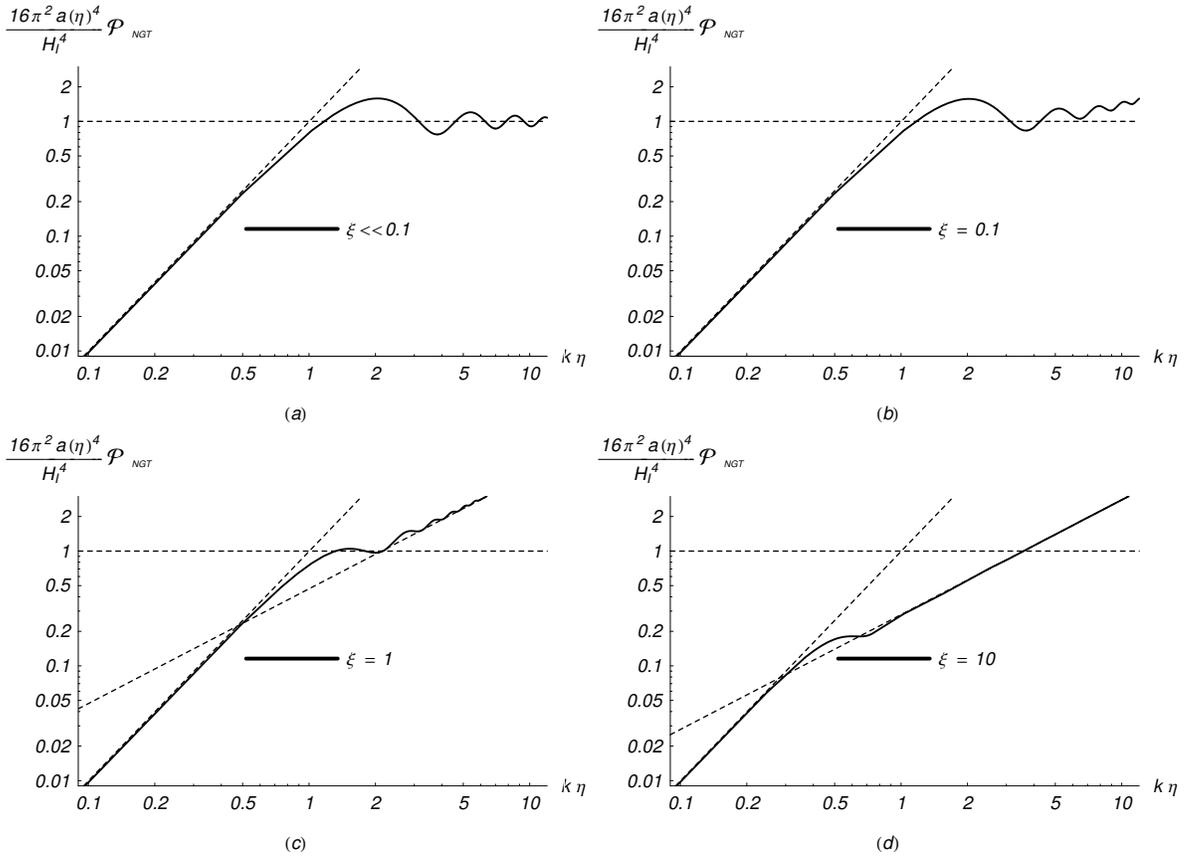, width=18cm}\caption{The
evolution of a specific mode in momentum space, plotted for several
values of $m_B H_I / k^2$. For low mass, a radiative spectrum is
retrieved which redshifts $\propto a^{-4}$. The larger the mass, the
earlier the evolution of a mode becomes massive as the mass exceeds
the Hubble factor, where the redshift becomes $\propto a^{-3}\propto
\eta\times a^{-4}$. }\label{fig:spec-rad-k}
\end{figure*}

\begin{figure*}[t]
\epsfig{file=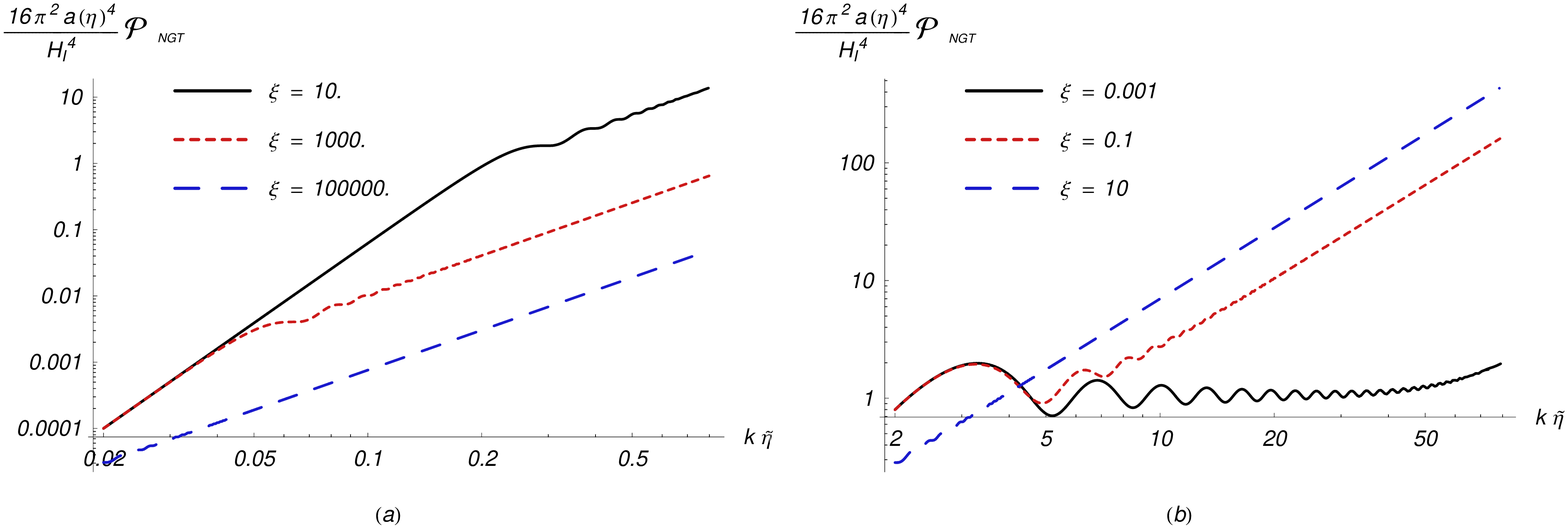, width=18cm} \caption{The
evolution of a specific mode in momentum space during the matter
era, plotted for several values of $m_B H_I / k^2$. For low mass, a
radiative spectrum is retrieved which redshifts $\propto a^{-4}$.
Again, the larger the mass, the earlier the evolution of a mode
becomes massive, with a redshift of $\propto a^{-3}\propto
\eta^2\times a^{-4}$. }\label{fig:spec-mat-k}
\end{figure*}

In both radiation and matter era the subhubble spectrum grows as
$\mathcal{P}_{\rm superH}(k)\propto k^2$, which is an infrared save
spectrum. In the ultraviolet region the spectrum converges to the
Bunch-Davies vacuum, $\mathcal{P}_{\rm subH}\propto
\mbox{constant}$. In the intermediate region, the mass determines
the height and blueness of the peak in the spectrum, hence in both
radiation and matter era the mass term causes an enhanced spectrum
with respect to the massless theory~\cite{Prokopec:2005fb}. The
evolution of one mode, fixed $k$, in both radiation and matter era
is given in figures~\ref{fig:spec-rad-k}~and~\ref{fig:spec-mat-k}.

\begin{figure}[t]
\epsfig{file=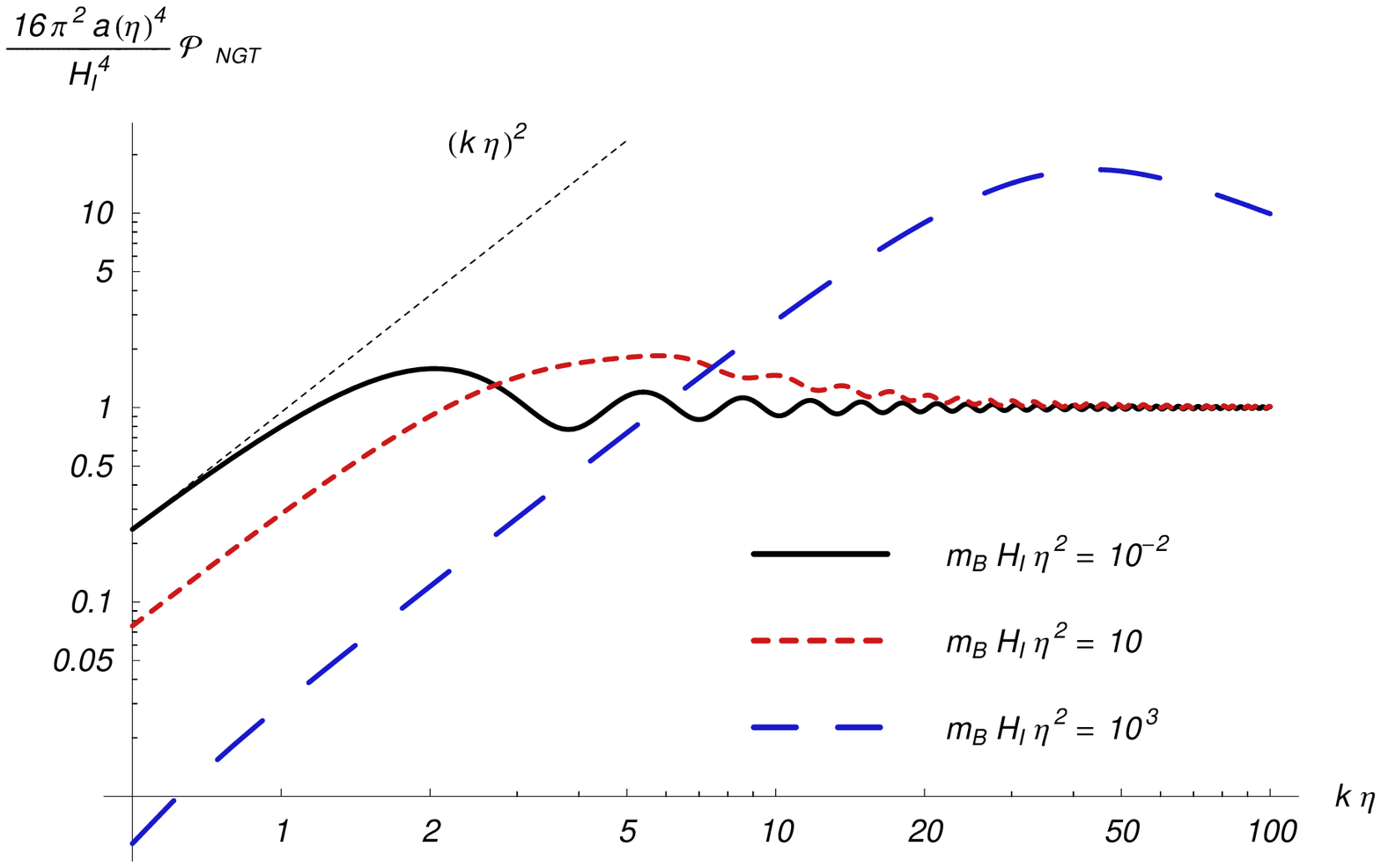, width=9cm} \caption{A
snapshot of the spectrum during the radiation era at a fixed moment
in time, plotted for several values of $m_B H \eta^2$. On
superhubble scales the spectrum grows as $k^2$, whereas in the
ultraviolet the spectrum becomes flat. Higher mass enhances the
spectrum and blueshifts the peak, which lies at $k\simeq\sqrt{m_B
H_I}$.}\label{fig:spec-rad-z}
\end{figure}

\begin{figure*}[t]
\epsfig{file=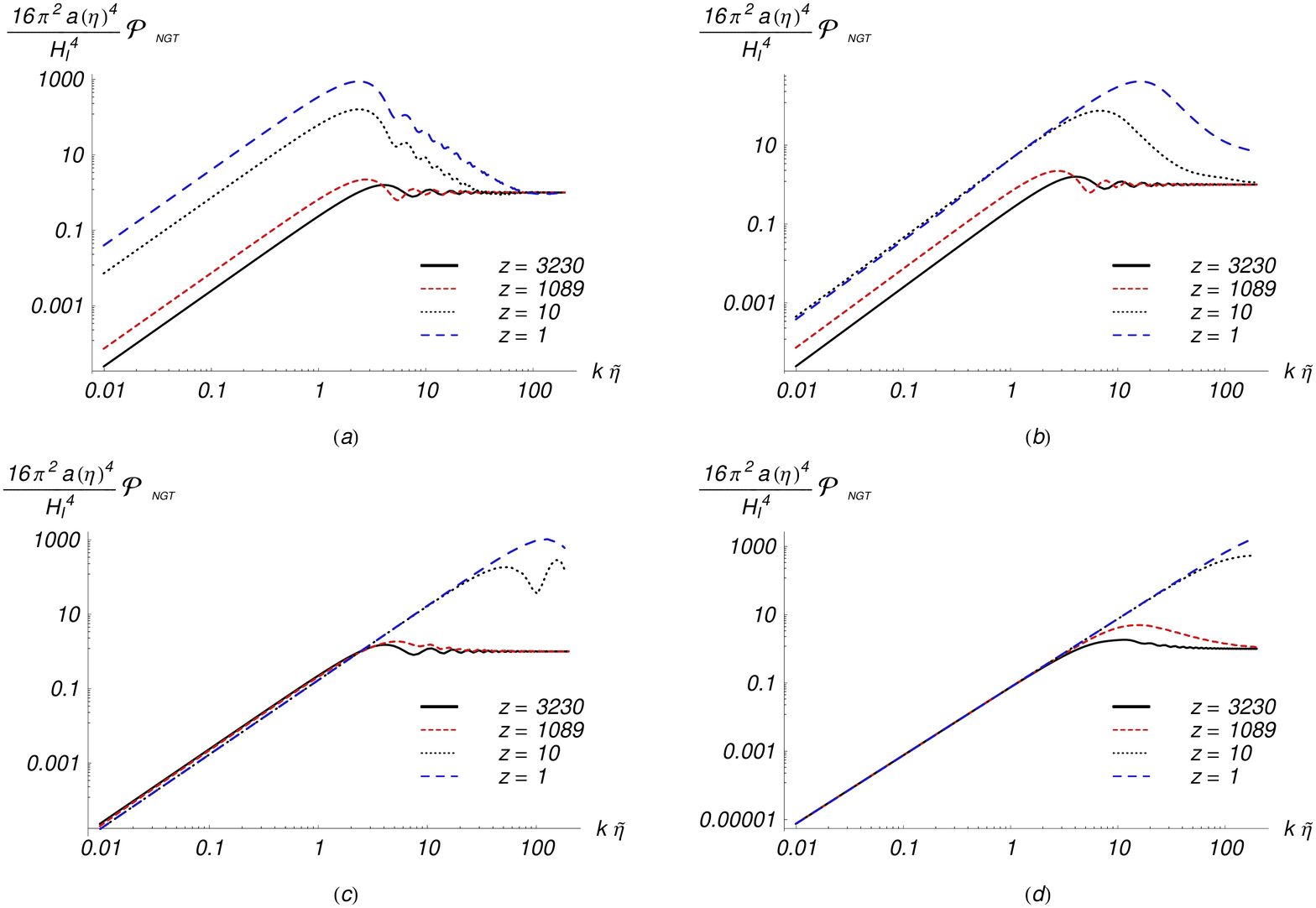, width=18cm}
\caption{Snapshots of the spectrum during the matter era, plotted
for several values of $\xi=m_B H \eta_e^2$. Figures (a)--(d)
represent $m_B H \eta_e^2 = 10^{-6}, 10^{-2}, 1$ and $10$,
respectively. The maximum value of the power spectrum still lies at
$k\simeq\sqrt{m_B H_I}$. }\label{fig:spec-mat-z}
\end{figure*}

Evolving in time, the damping of the modes in the density
fluctuations evolves as $\mathcal{P}\propto a^{-2}$ on superhubble
scales in both eras. For small enough mass, the spectrum remains
relativistic in subhubble regions, that is, a damping of
$\mathcal{P}\propto a^{-4}$. However, as soon the mass term $m_B^2
a(\eta){^2}$ becomes dominant in the field equation, the evolution
of a mode becomes non-relativistic, $\mathcal{P}\propto a^{-3}$,
with a decaying oscillatory behaviour on top. A snapshot of the
spectrum at fixed time $\eta$ is illustrated in
figures~\ref{fig:spec-rad-z}~and~\ref{fig:spec-mat-z}.

It is the aforementioned non-relativistic evolution,
depending on the scale factor, which makes the theory viable for
providing the energy density of dark matter.

\section{Energy density}
As the mass term in the field equations grows with the scale factor,
at a certain moment the energy density of the field switches from
being dominated by relativistic modes to domination by
non-relativistic modes, as previously explained.

When this switch in domination occurs, the energy density of the
field grows with respect to the total energy density of the universe
during the radiation dominated era.

If we assume that inflation can be described by a scalar
inflationary model, such that
\begin{align}
H_I^2=\frac{\rho_{\phi}}{3 M_P^2},
\end{align}
where $M_P=(8\pi G_N)^{-\frac{1}{2}}\simeq 2.4\times 10^{18}$ GeV is the
reduced Planck mass, the energy density of the radiation field at
the end of inflation equals
\begin{align}
\rho_{\rm rad} (\eta_{end})=3 H_I^2 M_P^2.
\end{align}

The energy density of the nonsymmetric metric field is given by
\begin{align}
\rho_{\rm B}(\eta)=\int\frac{dk}{k}{\cal P}_B(\eta,k).
\end{align}
In order to calculate the observable energy density, we 
subtract the conformal part of it, which effectively means 
that we impose an ultraviolet cut-off in the power spectrum. 
The excess energy-density comes from the modes 
which crossed the Hubble radius during inflation. 
The smallest of these momenta is given by the physical momentum
at the beginning of inflation, $k_{phys}|_{in} \equiv k/a(\eta_{in})=H_I$, 
while the largest momentum is given by the physical momentum
at the end of inflation, $k_{phys}|_{end} \equiv k/a(\eta_{end})=H_I$, 
where $\eta = \eta_{in}$ ($\eta = \eta_{end}$)
denotes the conformal time at the beginning (end) of inflation. 
The latter momentum is also taken as the ultraviolet cut-off.

As the variable over which is integrated is $k\eta$, the minimum and
maximum value of the domain of integration are given by $a_{in}
a(\eta)$ and $a_{end} a_{\eta}$ respectively. Thus the energy density of
the antisymmetric tensor field becomes
\begin{align}
\rho_{\rm B}=&\int_{a_{in}
H_I}^{a_{end} H_I} \frac{dk}{k}\mathcal{P}_{\rm B}\nonumber\\
%=&\int_{a_{in}a(\eta)}^{a_{end}a(\eta)}\frac{dk\eta}{k\eta}\frac{H_I^4}{4\pi^2a(\eta)^4}\left[\frac{1}{2(k\eta)^2}+1-\frac{\cos2k\eta}{2(k\eta)^2} - \frac{\sin 2k\eta}{k\eta}\right]\nonumber\\
=&\frac{H_I^4}{4 \pi^2 a(\eta)^4}\left[-\frac{1}{4x^2}+\log x +
%\right.\nonumber\\ &\left.
\frac{\cos 2x}{4x^2}+\frac{\sin 2x}{2x}-{\rm ci}
2x\right]_{a_{in} a(\eta)}^{a_{end} a(\eta)}.
\end{align}
where $x=k\eta$, $a_{in}=a(\eta_{in})$ and $a_{end}=a(\eta_{end})$. 
From equation~\eqref{eq:scalefactor-rad-era} we
have $a_{end}=1$, such that we may expand for small $a_{in}$,
\begin{align}
\rho_{\rm B}=&\frac{H_I^4}{4 \pi^2 a(\eta)^4}\left[\log a(\eta) -
\frac12 +\log2+\gamma_E
%\right.\nonumber\\&\left.
+\frac{1}{2} a_{in}^2
a(\eta)^2+\mathcal{O}\left(a(\eta)^{-2}\right)+\mathcal{O}\left(a_{in}^{3}\right)\right].\label{eq:energydensity-massless}
\end{align}
Here $\gamma_E$ denotes the Euler constant. Now we see that the
energy density contains a subhubble term scaling as $a(\eta)^{-4}$,
a term scaling as $a(\eta)^{-4} \log a(\eta)$ representing the
continuous filling up of the density on horizon scales, and a small
term $a_{in} a(\eta)^{-2}$ for the superhubble energy density. We
assume the scale of inflation to be large enough to have $a_{in}
a(\eta) < 1$, such that the expansion is indeed allowed.

When the energy density is dominated by non-relativistic modes, the
calculation of the energy density is less trivial. As the field is
given in terms of $M_{\frac{i k^2}{4m_B H_I},\pm \frac{3}{4}}(-i m_B
H_I \eta^2)$, the integral over all fourier modes too complicated to
be evaluated exactly in an analytical fashion.
However, we do know the asymptotic behaviour
of the field in the infrared and the ultraviolet regions separately.

In the infrared the leading order of the spectrum is $\propto
(k\eta)^2$. In the ultraviolet the spectrum becomes that of the
massless theory.

This information together with a closer look at the field equation,
\begin{align}
\left(\partial_0^2+k^2-\frac{2}{\eta^2}+m_B^2H_I^2\eta^2\right)\tilde{B}^L_{\rm
rad}(\eta,k)=0.\label{eq:eom-bl-rad2}
\end{align}
allows a simple approximation. Note that during the radiation era
the term $\propto\eta^{-2}$ represents $a(\eta)^2 H(\eta)^2$, where
$H(\eta)$ is the Hubble factor.

The approximation is the following: each of the second, third and
fourth term in equation~\eqref{eq:eom-bl-rad2} has its own region in
which it defines the behaviour of the field, but for each value of
$k$ the mass eventually will dominate, leading to a scaling of
$a(\eta)^{-3}$. Thus the energy density during the period of massive
behaviour can be given in terms of the energy density during the
period of massless behaviour.

The superhorizon modes all together are the first ones to start
scaling as non-relativistic matter, as soon as the Hubble constant
shrinks about below the mass, $m_B > \sqrt{2} H(\eta)$. That is the
case when $a(\eta)\simeq \sqrt{H_I / m_B}$, such that
\begin{align}
\mathcal{P}_{\rm superH}(k,\eta)\simeq\mathcal{P}^{\rm
massless}_{\rm superH}(k,\eta) \rightarrow a(\eta)
\sqrt{\frac{m_B}{H_I}}\mathcal{P}^{\rm massless}_{\rm
superH}(k,\eta) .
\end{align}
For sub-Hubble modes, the field starts scaling massive as soon as
the wave vector shrinks below the mass, $m_B > k / a(\eta)$, when
$a(\eta)\simeq k / m_B$, such that
\begin{align}
\mathcal{P}_{\rm subH}(k,\eta)\simeq\mathcal{P}^{\rm massless}_{\rm
subH}(k,\eta) \rightarrow \frac{m_B\, a(\eta)}{k}\mathcal{P}^{\rm
massless}_{\rm subH}(k,\eta)
\end{align}
$\mbox{ for $H(\eta) < \frac{k}{a(\eta)} < m_B$}$. The peak of the
massive spectrum lies at the mode of which the physical momentum is
equal to the mass when it crosses inside the Hubble radius, that is, $k^2
= a(\eta)^2 H(\eta)^2 = a(\eta)^2 m_B^2$,
\begin{align}
k_{peak}=&m_B\, a(\eta_{NR})\nonumber\\
=&\sqrt{H_I m_B},
\end{align}
where $\eta_{NR}$ denotes the moment in time of switching from
relativistic to non-relativistic for the mode $k_{peak}$.

Now we can approximate the non-relativistic energy density by
\begin{align}
\rho_{\rm B}
 =&  \int_{a_{in}
H_I}^{\sqrt{m_B H_I}} \frac{dk}{k} a(\eta) \sqrt{\frac{m_B}{H_I}}
\mathcal{P}^{\rm massless}_{\rm superH}(k,\eta)
% \nonumber\\ &
+ \int_{\sqrt{m_B H_I}}^{m_B a(\eta)} {dk}\frac{m_B \,a(\eta)}{k^2}
\mathcal{P}^{\rm massless}_{\rm subH}(k,\eta)
\nonumber\\& 
+ \int_{m_B a(\eta)}^{a_E H_I} \frac{dk}{k} \mathcal{P}^{\rm
massless}_{\rm subH}(k,\eta) \nonumber\\
=&\, \frac{H_I^4}{8 \pi^2 a(\eta)^3}\sqrt{\frac{m_B}{H_I}}
\left[\mathcal{B} -2a(\eta)^{-1}\log \frac{m_B\,a(\eta)}{H_I}- 2
\sqrt{\frac{m_B}{H_I}}\right.\nonumber\\
&\left.+\mathcal{O}\left(a_{in}^2a(\eta)^2,a(\eta)^{-2},(\sqrt{m_B
H_I}\eta)^{-2}\right)\right].
\end{align}
In this expansion, $\mathcal{B}$ is the leading order coefficient, which can be
determined by a numerical calculation of the energy density. In our
approximation with a sudden relativistic-to-non-relativistic
transition, we find $\mathcal{B}\simeq
-2^{\frac{1}{4}}+2+2^{\frac34}\log 2 + 2^{\frac34}\gamma_E\simeq
1.35$. The third integral term only contributes if $m_B a(\eta) <
H_I$. Of course, when $m_B a(\eta) > H_I$, the upper bound of the
second integral becomes $a_E H_I$, and the logarithmic term in the
energy density becomes nonexistent. This logarithmic term represents
the contribution of superhubble but relativistic modes. This term
damps quickly, and is neglected in the following. We assume that the
scale of inflation is large enough to let $a_{in} a(\eta)\ll 1$
always, and we already made use of the fact that during the
non-relativistic behaviour of the theory $\sqrt{m_B H_I}\eta
> 1$.

Taking account only for the leading order, we find for the
NGT-to-radiation ratio when the field has become massive,
\begin{align}
\frac{\rho_{\rm B}}{\rho_{\rm rad}}=&\epsilon \, a(\eta)
\sqrt{\frac{m_B}{H_I}}\qquad {\rm for\,\,}\epsilon = \frac{H_I^2
\mathcal{B}}{3 \pi m_P^2},
\end{align}
where $m_P^2=8 \pi M_P^2$ is the unreduced Planck mass.

In order for the energy density to fully take account for the
dark-matter energy density, it has to satisfy
\begin{align}
\frac{\rho_{\rm B}}{\rho_{\rm rad}}=1\qquad\mbox{at $\eta=\eta_e$}.
\end{align}
Hence,
\begin{align}
\epsilon \, {a_{\rm eq}}\sqrt{\frac{m_B}{H_I}}=1
\end{align}
This gives a rough approximation for the mass,
\begin{align}
m_B&=\frac{H_I}{\epsilon^2 a_{\rm eq}^2}
= \frac{H_{\rm eq}}{\epsilon^2}\nonumber\\
&=2.8\times10^{-2}\left(\frac{10^{13} \,\mbox{GeV}}{H_I}\right)^4\, \mbox{eV}
\,,
\label{m_B}
\end{align}
or
\begin{align}
m_B^{-1}&\simeq 7\times10^{-8}
\left(\frac{H_I}{10^{13}\,\mbox{GeV}}\right)^4\, \mbox{m}.
\end{align}
In the last line we used for the Hubble parameter at radiation-matter
equality, $H_{\rm eq}=(z_{\rm eq}+1)^{3/2}H_0$, $H_0\simeq
1.4\times10^{-42}\mbox{ GeV}$, $z_{\rm eq}=3230\pm200$ and the scale
factor $a_{\rm eq}/a_0=1/(z_{\rm eq}+1)$.

\section{Pressure}
Similar to the energy density, the pressure of the theory can be
calculated as the pressure is given by the spatial components of the
diagonal of the stress-energy tensor,
\begin{align}
T_i\/^i = - 3 P.
\end{align}
If isotropy is assumed, that is $k_ik^j = \tfrac13 \delta_i\/^j
\left| k\right|^2$ for $i=j$, the expression for the 'pressure per
mode' $P$ becomes,
\begin{align}
P(\eta,k)=&\frac{-1}{24\pi^2 a(\eta)^{4}}\left[\left|\partial_0
\tilde{B}^L(\eta,k)+\frac{a(\eta)'}{a(\eta)}\tilde{B}^L(\eta,k)\right|^2\right.\nonumber\\
&\left.-3k^2 \left|\tilde{B}^L(\eta,k)\right|^2-m_B^2
a(\eta)^2\left|\tilde{B}^L(\eta,k)\right|^2\right].\label{eq:pressure-per-mode}
\end{align}
The physical pressure can then be calculated by
\begin{align}
P(\eta)=\int\frac{dk}{k}P(\eta,k).
\end{align}
When the field is dominated by relativistic modes, during radiation
era this leads to,
\begin{align}
P(\eta)&=\frac{1}{4\pi^2 a(\eta)^{4}}\left[\frac13 \log
a(\eta)+\frac13 \log 2 + \frac13 \gamma_E-\frac12\right.\nonumber\\
&\left. - \frac{1}{6} a_{in}^2 a(\eta)^2 + \frac{\sin
2a(\eta)}{3a(\eta)} +\mathcal{O}\left(a_{in}^4
a(\eta)^4,a(\eta)^{-2}\right)\right].\label{eq:pressure-massless}
\end{align}
Covariant conservation of the stress-energy tensor implies
\begin{align}
\rho_B'+3\frac{a(\eta)'}{a(\eta)}\left(\rho_B+P\right)=0.
\end{align}
Comparing
expressions~\eqref{eq:energydensity-massless}~and~\eqref{eq:pressure-massless},
we find that energy conservation is indeed obeyed order by order in
$\eta$.

On superhubble scales, $k\eta\ll1$, the power spectrum and the
pressure per mode are dominated by the first terms in
relations~\eqref{eq:powerspectrum}~and~\eqref{eq:pressure-per-mode}.
Thus one finds that for superhubble modes, both relativistic and
non-relativistic, we have $P(\eta,k)\simeq -\tfrac13
\mathcal{P}_{\rm B}(\eta,k)$, in accordance with the scaling of
$\rho \propto a(\eta)^{-2}$.

During inflation and during the matter era similar results are found
for the relativistic modes, satisfying energy conservation.

This integration must be performed for the massive modes as well.
However, for the same reasons as for the energy density, the
pressure for the massive modes can only be either approximated or
integrated numerically. We do not perform here a complete analysis
of the pressure. Instead, we illustrate the pressure for one
specific choice of parameters, and comment on how the pressure
behaves for a general choice of parameters.

\begin{figure}[t]
\epsfig{file=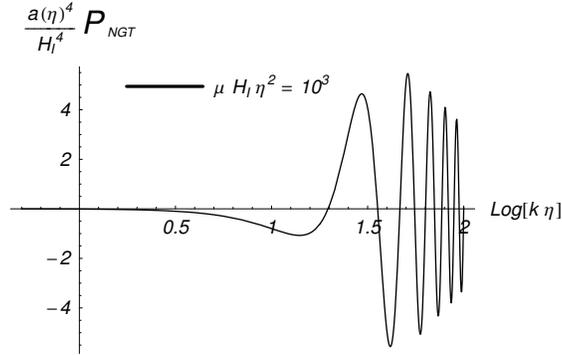, width=8cm} \caption{The 'pressure
per mode' for non-relativistic modes during radiation era, for $m_B
H_I \eta^2 = 10^3$. Note that the x-axis is logarithmic, where the
y-axis is linear. For superhubble scales the pressure scales as
$-\tfrac13 \rho$, whereas on subhubble scales is averages aroung
zero.}\label{fig:pressure}
\end{figure}

In figure~\ref{fig:pressure} the 'pressure per mode' is illustrated
for a massive nonsymmetric field, with $m_B H_I \eta^2 = 10^3$. On
superhubble scales, that is $k\eta\ll1$, the pressure per mode
behaves as mentioned above as $P(\eta,k) = - \tfrac13
\mathcal{P}_{\rm B}(\eta,k)$. On subhubble scales the pressure per
mode becomes zero, conserving energy with a density scaling as $\rho
\propto a(\eta)^{-3}$.

During inflation and matter era the behaviour of the pressure is
consistent with covariant energy conservation as well. The result is
that we have zero pressure at the time of decoupling when the field
is massive. This is in agreement with the requirements on cold dark
matter.

\section{Discussion and conclusions}

 We have considered the cosmology of a massive antisymmetric tensor field
whose dynamics is given by the action~(\ref{general lagrangian}).
We follow the evolution
of the vacuum fluctuations generated during an inflationary epoch,
through inflation, radiation and matter era.
We find that the antisymmetric field with a mass,
$m_B=3\times10^{-2}\left({10^{13} \,\mbox{GeV}}/{H_I}\right)^4\, \mbox{eV}$,
results in the right energy density today to account for the dark matter
of the Universe.
This then implies that below the corresponding length scale,
$m_B^{-1} \simeq 0.1\,\,{\rm \mu m}
   \left({H_I}/{10^{13} \,\mbox{GeV}}\right)^4$,
the strength of the gravitational coupling may change,
as it has been argued in Refs.~\cite{Moffat:1995dq,Moffat:2004bm}.
This scale is about two orders of magnitude below the present
experimental bound, which is of the order
$10 \mu\mbox{m}$~\cite{Will:2005va}. Note that the mass scale~(\ref{m_B})
depends strongly on the scale of inflation, such that,
if the gravitational force law below the scale of
$0.1\,\,{\rm \mu m}$ remains unchanged,
would imply that, either the $B$ field is nondynamical, or
the inflationary scale is lower than $10^{13}~{\rm GeV}$.
Note that, in contrast to gravitons, the antisymmetric tensor field
begins scaling as non-relativistic matter during radiation era, making it
potentially the most sensitive probe of the inflationary scale.

The peak in the energy density power spectrum, generated as a
consequence of a breakdown of conformal invariance in radiation era,
corresponds to a comoving momentum scale of the order, $k\simeq
\sqrt{m_B H_0}/(1+z_{\rm eq})^{1/4}$, where $z_{\rm eq}\simeq 3230$
is the redshift at matter-radiation equality.
 Hence the most prominent matter density perturbations at a redshift $z=10$,
when structure formation begins,
occur at a scale, $k_{\rm phys}^{-1}\sim 2\times 10^{7}\,{\rm km}$
(corresponding to a wavelength, $\lambda_{\rm phys}
\sim 1\times 10^{8}~{\rm km}$, which is (co-)incidentally
the Earth-Sun distance),
which may boost early structure formation on these scales.
We therefore expect that, when compared to other CDM models,
this type of cold dark matter may induce an earlier structure formation.
Even though the mass of the antisymmetric field is quite small
(of the order of the heaviest neutrino mass,
$m_{\nu} \sim 0.1\,\,{\rm eV}$), the antisymmetric tensor field dark matter
is neither hot nor warm. Indeed, since the antisymmetric tensor field does
not couple to matter fields,
it cannot thermalise and its spectrum remains primordial, and thus
highly non-thermal. Hence, in spite of its small mass the field pressure
remains small, such that structure formation on small scales
does not get washed out. Indeed, 
$P/\rho_B \simeq E_{kin}/(m_B c^2) \sim 10^{-34}$
evaluated at ${k_{phys}=(k_{peak}/a)|_{today}}$, which is a tiny pressure.

Although the pressure converges to zero as the field becomes
dominated by non-relativistic modes, the pressure per mode in Fourier
space shows a characteristic oscillatory behaviour shown
 in figure~\ref{fig:pressure}. These oscillations
may influence cosmological perturbations in a manner
analogous to Sakharov oscillations~\cite{Albrecht:1992kf}, which in turn
may produce an observable imprint in the cosmic background photon fluid.
This question deserves further investigation.

%GATHER{refs.bib}
\bibliography{refs}
\end{document}